\documentclass[aps,twocolumn,prl]{revtex4-2}
\usepackage{amsmath}
\usepackage{amssymb}
\usepackage{mathdots} 
\usepackage{cancel} 
\usepackage{braket} 
\usepackage{wrapfig} 
\usepackage{setspace} 
\usepackage[breaklinks,pdfborder={0 0 0},colorlinks=true,linkcolor=blue, citecolor=blue, urlcolor=blue]{hyperref}
\usepackage{lipsum}
\usepackage{graphicx}
\usepackage{dcolumn}
\usepackage{bm}
\usepackage{lmodern}
\usepackage{fix-cm}
\usepackage[mathscr]{eucal} 

\counterwithin*{figure}{section}
\counterwithin*{figure}{subsection}

\usepackage{xcolor}
\makeatletter
\def\@hangfrom@section#1#2#3{\@hangfrom{#1#2}#3}%
\def\@hangfroms@section#1#2{#1#2}%
\makeatother
\newcommand{\dr}{\mathrm{d}^d r}
\newcommand{\dR}{\mathrm{d}^d R}
\newcommand{\dk}{\mathrm{d}^d k}
\newcommand{\dq}{\mathrm{d}^d q}
\newcommand{\e}{e}
\newcommand{\bracket}[2]{\langle #1 | #2 \rangle}
\newcommand{\dd}{\mathrm{d}}
\newcommand{\bA}{\mathbf{A}}
\newcommand{\bR}{\mathbf{R}}
\newcommand{\br}{\mathbf{r}}
\newcommand{\bk}{\mathbf{k}}
\newcommand{\bG}{\mathbf{G}}
\newcommand{\bq}{\mathbf{q}}
\newcommand{\bkappa}{\boldsymbol{\kappa}}
\newcommand{\Tr}{\mathrm{Tr}}
\newcommand{\tr}{\mathrm{tr}}

\makeatletter
\AtBeginDocument{\let\LS@rot\@undefined}
\makeatother

\begin{document}

\raggedbottom

\title{Magneto-ARPES as a direct probe of Berry Curvature and the Orbital Magnetic Moment}
\author{Seohyun Kong}
\affiliation{SUPA, School of Physics and Astronomy, University of St Andrews, North Haugh, St Andrews, KY16 9SS, United Kingdom}
\author{Bernd Braunecker}
\affiliation{SUPA, School of Physics and Astronomy, University of St Andrews, North Haugh, St Andrews, KY16 9SS, United Kingdom}
\author{Luke C. Rhodes}
\email[Correspondence to: ]{lcr23@st-andrews.ac.uk}
\affiliation{SUPA, School of Physics and Astronomy, University of St Andrews, North Haugh, St Andrews, KY16 9SS, United Kingdom}

\begin{abstract}
The Berry curvature and orbital magnetic moment are central to the understanding of electronic transport phenomena such as the anomalous Hall effect, yet measuring these quantities directly with momentum resolution remains experimentally challenging. Here we show that angle-resolved photoemission spectroscopy measurements in a weak magnetic field (magneto-ARPES) provide a direct route to both quantities. We develop an analytic theory of photoemission in a weak magnetic field and show that, for any material, the first-order change in photoemission intensity is directly proportional to both the Berry curvature and the orbital magnetic moment, even in the presence of self-energy effects. For linearly polarised light, the matrix elements are unchanged to first order in the field, so the intensity change directly reflects these quantities. Using monolayer WSe$_2$ as an example, we show how both quantities can be extracted in a typical experiment. This finding thus points to magneto-ARPES as a general route for measuring the momentum-resolved Berry curvature and orbital magnetic moment in quantum materials.
\end{abstract}

\maketitle

\pagenumbering{arabic} 

The Berry curvature and orbital magnetic moment (OMM) are two cornerstones of condensed matter theory. They are known to directly govern the anomalous Hall effect \cite{TKNN1982}, orbital Hall effect \cite{Choi2023} and low-field magnetoresistance \cite{THONHAUSER2011}. The Berry curvature is a central pillar of the development of topological materials, due to its quantized nature when integrated over the Brillouin zone in non-interacting 2D materials \cite{Barry1983,Berry1984}. Considerable effort has been spent cataloguing topological phases in real materials, with high-throughput schemes identifying thousands of candidate materials from first-principles calculations \cite{Vergniory2022}. However, despite this progress, systematic direct measurement of the momentum-resolved Berry curvature and OMM remains experimentally challenging.

Angle-resolved photoemission spectroscopy (ARPES) directly resolves the momentum-space electronic structure of quantum materials, and has been central to identifying topologically protected surface states in topological insulators \cite{Chen2009_Bi2Te3, Hsieh2009}. It has previously been shown that the use of circularly polarized light can produce an asymmetry of the matrix elements of certain bands in such a way as to be able to extract information about the pseudospin and the ${\bf d}$-vectors of suitable bands, which can be related back to the Berry curvature, OMM, and the quantum geometric tensor \cite{Wang2013, Rostami2019, Cho2018, Cho2021, Kim2025}. However, this approach is accompanied by the experimental challenge of disentangling matrix element asymmetry from the alignment of the photon beam and sample geometry, and has been limited to a subset of Dirac-like band dispersions.

Here, we propose an alternative experiment utilizing ARPES that can enable the direct extraction of the Berry curvature and OMM for any material via the application of a weak magnetic field. Magnetic fields within the ARPES apparatus have traditionally been avoided, since the Lorentz force distorts photoelectron trajectories and degrades momentum resolution. However recent works \cite{Ryu2023, Huang2023, Huang2026,ObandoGuevara2026} have overcome this limitation, showing that the spectral function can still be measured under fields of $10-100$~mT once the resulting photoelectron deflection has been systematically corrected for, a technique dubbed magneto-ARPES.

In this paper, we show that the first order correction to the photoemission intensity under a weak magnetic field, is directly proportional to both the Berry curvature and the OMM, and we show how these quantities can be extracted by measuring the difference in photoemission signal, even in the presence of self-energy effects. We demonstrate that linearly polarized light does not affect the matrix elements to first order. This work thus establishes magneto-ARPES as a route to the direct, momentum-resolved measurement of both the Berry curvature and the orbital magnetic moment, opening the geometry of the electronic wavefunction for direct experimental study.

ARPES measures the photoelectron intensity $I ({\bf k}, \omega)$ as a function of energy $\omega$ and momentum ${\bf k}$, which, under the three-step model and the sudden approximation, is conventionally written as \cite{Damascelli2003, Moser_2017, King2020}
\begin{equation} \label{eq:Photoemission_intensity}
        I({\bf k}, \omega) 
        = \frac{2\pi}{\hbar}\sum_{f, i} | M_{f, i} ({\bf k}) |^{2} \, A ({\bf k},\omega) \, f (\omega) 
        \text{.}
\end{equation}
Here $|M_{f, i} ({\bf k}) |^{2}$ are single particle matrix elements encoding the probability of an excitation of an electron by a photon from an initial state $i$ to a final state $f$. $A ({\bf k}, \omega)$ is the single particle spectral function, and $ f (\omega) $ is the Fermi-Dirac distribution. 

Understanding how Eq.~\eqref{eq:Photoemission_intensity} is modified in a weak magnetic field requires care, since the field changes the meaning of the momentum variable. Minimal coupling replaces the canonical wavevector $\hat{\bm{\kappa}}$ by the kinetic wavevector $\hat{\mathbf{k}} = \hat{\bm{\kappa}} + \tfrac{e}{\hbar} {\bf A} (\hat{\bf r})$, with ${\bf A}$ the vector potential and $-e$ the electron charge. Assuming the magnetic field $B_z$ is applied along the $z$ direction, the $(x,y)$ components of $\hat{\bf k}$ then no longer commute, $[\hat{k}_{x}, \hat{k}_{y}] = - i e B_{z} / \hbar$, so momentum-resolved quantities cannot be multiplied as ordinary functions. Moreover, the vector potential is not translationally invariant and real-space kernels $\braket{{\bf r}_{2} | \hat{O} | {\bf r}_{1}}$ of any operator $\hat{O}$ acquire gauge-dependent phases and their subsequent Fourier transform is no longer gauge invariant. However, it is still possible to write the operator in a mixed real and momentum space basis by defining a coordinate substitution of $\mathbf{R}=\frac{1}{2}({\bf r_1}+{\bf r_2})$ and a relative coordinate $\mathbf{r} = {\bf r_2}-{\bf r_1}$ and Fourier transforming over the relative coordinate \cite{Moyal_1949,Hillery1984,Onoda2006Keldysh, Sugimoto2007GaugeCovariant}, in $d$ dimensional space,
\begin{equation} \label{Eq:Weyl}
    O_{W} ({\bf R}, {\bf k}) = \int \mathrm{d}^{d} r \, 
    e^{-i {\bf k} \cdot {\bf r}}e^{- i \Phi}
    \langle {\bf R} + \tfrac{{\bf r}}{2} | \hat{O} | {\bf R} - \tfrac{{\bf r}}{2} \rangle
    \text{,}
\end{equation}
where $O_{W}$ is known as the Weyl symbol of the operator and the phase $\Phi ({\bf R}, {\bf r}) = - \tfrac{e}{\hbar} \int_{{\bf R} - {\bf r}/2}^{{\bf R} + {\bf r}/2} \mathrm{d} \bm{\ell} \cdot {\bf A} (\bm{\ell})$ enters through the Wilson line $e^{- i \Phi}$ that cancels the gauge-dependent phase of the kernel \cite{SM}. For products of operators, the Fourier transform acquires additional terms due to the noncommutativity of position and momentum. These terms are encoded within the Moyal star product \cite{Hillery1984} $(\hat{f} \hat{g})_{W} = f_{W} \star g_{W}$ \cite{SM}. For the case of a magnetic field along $B_z$ this star operator can be decomposed into two factors $\star = \star_0\star_B$ \cite{Polkovnikov2010, Onoda2006Keldysh, Sugimoto2007GaugeCovariant}, where
\begin{align}
    &\star_0 = 
    \exp \left( \frac{i}{2}
        \left[
            \overleftarrow{\nabla}_{\bf R} \cdot \overrightarrow{\nabla}_{\bf k} 
            - \overleftarrow{\nabla}_{\bf k} \cdot \overrightarrow{\nabla}_{\bf R} 
        \right]
    \right),
 \label{Eq:star_zero}
 \\
    &\star_B =\exp \left(
    - \frac{i e B_{z}}{2 \hbar}
        \left[
            \overleftarrow{\nabla}_{\bf k} \times \overrightarrow{\nabla}_{\bf k} 
        \right]_{z}
    \right),
 \label{Eq:star_B}
\end{align}
and the arrows above the derivative operators $(\nabla_i)$ define whether the derivative acts on the Weyl symbol on its left or right. It is possible to rewrite Eq.~\eqref{eq:Photoemission_intensity} in terms of Weyl symbols and star products, such that \cite{SM}

\begin{equation} \label{Eq:PE_Moyal_grouped}
    \begin{split}
        I ({\bf k}_{f}, &\omega) = \frac{2 \pi}{\hbar} \sum_{n, m, \chi}
        \int \frac{\mathrm{d}^{d} R \, \mathrm{d}^{d} k}{(2 \pi)^{d}} \, f (\omega) \\
        &\times \mathrm{tr} \bigl\{
            M_W \star A^{(n,m)}_{W} \star M^{*}_W \star  \big(| \chi, {\bf k}_{f} \rangle\bigr\langle \chi, {\bf k}_{f}|\big)_W 
        \bigr\}
        \text{,}
    \end{split}
\end{equation}
with $\mathrm{tr}$ the trace over the spin degrees of freedom. $M_W$ is the Weyl symbol of the photoemission dipole operator $\hat{M}=\tfrac{e \hbar}{m} {\bf A}_{\text{ph}}\cdot\hat{\mathbf{k}}$, with ${\bf A}_{\text{ph}}$ the photon vector potential \cite{Damascelli2003} that, in the dipole approximation, is independent of ${\bf R}$. $A_W$ is the Weyl symbol of the spectral operator, as obtained from the Green's function, and $\big(| \chi, {\bf k}_{f} \rangle\bigr\langle \chi, {\bf k}_{f}|\big)_W $ is the Weyl symbol of the projector to the final state, with kinetic 
$\mathbf{k}_f$ and spin state $\chi$.

We can determine the first order correction to the photoemission intensity by expanding Eq. \eqref{Eq:PE_Moyal_grouped} in powers of $B_z$. This involves both the expansion of the star product as well as the expansion of the Weyl symbols $O_W = O^{(0)} + O^{(1)} + ... $ where $O^{(1)} \propto B_z$. The zeroth order term reproduces the conventional photoemission expression of Eq.~\eqref{eq:Photoemission_intensity}, now denoted as 
\begin{equation} 
\label{eq:Photoemission_intensity_zerothOrder_Linear}
    I^{(0)}({\bf k}, \omega) = 
    \frac{2 \pi}{\hbar} |M^{(0)} ({\bf k})|^2 \, A^{(0)} ({\bf k}, \omega) \, f(\omega)
    \text{.}
\end{equation}
The first order correction in $B_z$, contains three terms \cite{SM}, two of these terms depend on $\text{Im}[\mathbf{A}_{\text{ph}} \times \mathbf{A}_{\text{ph}}^*]$, which are only non-zero if $\mathbf{A}_{\text{ph}}$ is complex. These are important for the photoemission intensity measured using circularly polarized light, but do not impact the photoemission intensity measured with linearly polarized light, where $\mathbf{A}_{\text{ph}}$ can be chosen as real. Restricting henceforth to the case of linearly polarized light, the remaining first order correction is given by 
\begin{equation} \label{eq:Photoemission_intensity_1stOrder_Linear}
    I^{(1)}({\bf k}, \omega; B_z) = 
    \frac{2 \pi}{\hbar} |M^{(0)} ({\bf k})|^2 \, A^{(1)} ({\bf k}, \omega; B_{z}) \, f(\omega)
    \text{.}
\end{equation}
We emphasize that here the matrix elements $M^{(0)}(\mathbf{k})$ are the same as at zero field, so that the correction to the photoemission intensity under a magnetic field with linearly polarized light is entirely due to the field induced modification to the spectral function. 

The correction $A^{(1)}$ is obtained through the fundamental relation $A^{(1)} = - \tfrac{1}{\pi} \mathrm{Im} \, \mathrm{Tr} \, G^{(1)}$ where $G^{(1)}$ is the first-order correction to the retarded Green's function. The latter is computed through the Dyson equation, whose Weyl-space form is given by 
\begin{equation} \label{eq:MagneticMoyalProduct}
    (\omega - H_W - \Sigma_W) \star G_W 
     = 1
    \text{,}
\end{equation}
where $H_{W}$, $\Sigma_{W}$, and $G_{W}$ denote the Weyl symbols of the Hamiltonian, the self-energy, and the Green's function respectively. Expanding the star product and Weyl symbols \cite{SM} and transforming back from Weyl symbols, we find that $G^{(1)}$ splits into a local and an itinerant part,
\begin{equation} \label{eq:FirstOrderCorrection}
    \begin{split}
        G^{(1)} ({\bf k}, \omega; B_z) 
        = \underbrace{G^{(0)} (H^{(1)} + \Sigma^{(1)}) G^{(0)}}_{\text{local}} 
        \qquad \\
        + \underbrace{\frac{i e \hbar B_{z}}{2} G^{(0)} (\tilde{v}_{y} G^{(0)} \tilde{v}_{x} - \tilde{v}_{x} G^{(0)} \tilde{v}_{y}) G^{(0)}}_{\text{itinerant}}
        \text{,}
    \end{split}
\end{equation}
where $\partial_{k_{\mu}} ({H^{(0)}+\Sigma^{(0)}}) = \hbar \tilde{v}_{\mu}$ is the renormalized band velocity and $\partial_{k_{\mu}} G^{(0)}= \hbar G^{(0)} \tilde{v}_{\mu} G^{(0)}$. Here $H^{(1)} = \mu_{B} B_{z} (L_{z} + g_{s} S_{z})$ is the standard first-order perturbation theory result for a Bloch Hamiltonian in a magnetic field, with $L_{z}$ and $S_{z}$ being orbital and spin angular momentum operators, $\mu_{B}$ the Bohr magneton, and $g_{s}$ the electron spin $g$-factor, which combined give the conventional Zeeman splitting. $\Sigma^{(1)}$ depends on the self-energy being considered, but nevertheless will act to renormalize the bare orbital and Zeeman terms in $H^{(1)}$. Equation ~\eqref{eq:FirstOrderCorrection} relates the Green's function in a weak magnetic field directly to its zero-field counterpart and can be calculated numerically for any electronic structure and general self-energy function.

To extract analytical insight, however, we take the self-energy to be band-diagonal, so that the zero-field Green's function can be written as a resolvent
\begin{equation}\label{eq:resolvent}
    G^{(0)} ({\bf k}, \omega) = \sum_{n} \frac{\ket{n^{(0)}} \bra{n^{(0)}}}{\omega - E_{n}^{(0)}}
    \text{,} \quad
    E_{n}^{(0)} = \epsilon_{n}^{(0)} + \Sigma_{n}^{(0)}
    \text{,}
\end{equation}
with $\epsilon_{n}^{(0)}$ the $n$-th eigenvalue of the zero-field non-interacting Hamiltonian
and projectors $\ket{n^{(0)}}\bra{n^{(0)}}$. This approximation is exact when the orbitals forming the bands share the same self-energy, as is often enforced by symmetry. More generally, mixing of bands by the self-energy requires both a difference between orbital self-energies and a hybridization between those orbitals, and is suppressed by the energy separation of the bands, so it becomes important only near crossings of bands of different orbital character \cite{SM}. Substituting Eq.~\eqref{eq:resolvent} into Eq.~\eqref{eq:FirstOrderCorrection} and taking the trace via $\text{Tr} \; G^{(1)} = \sum_{m} \braket{m^{(0)}| G^{(1)}| m^{(0)}}$ \cite{SM} gives
\begin{equation} \label{Eq:GB_Trace}
    \begin{split}
        \mathrm{Tr} \, G^{(1)} ({\bf k}, \omega; B_{z}) = &\frac{e B_{z}}{\hbar} \sum_{n} \frac{\Omega_{n}^{z} }{\omega - E_{n}^{(0)} } \\ 
        - &B_{z} \sum_{n} \frac{\tilde{m}_{n}^{z} }{(\omega - E_{n}^{(0)} )^{2}}
    \end{split}
\end{equation}
where 
\begin{equation} \label{eq:BerryCurvature}
    \Omega_{n}^{z} ({\bf k}) =
    - 2 \hbar^{2} \sum_{m \neq n}
    \frac{\mathrm{Im} (v_{x}^{(n, m)} v_{y}^{(m, n)})}{(\epsilon_{n}^{(0)} - \epsilon_{m}^{(0)})^{2}}
    \text{,}
\end{equation}
is the TKNN formula for the bare Berry curvature of band $n$ along $z$ \cite{TKNN1982, Xia02010}. Note that $\Omega_{n}^{z}$ is along the $z$ axis and is built from bare eigenvalues and bare velocities $v_{x,y}^{(n,m)} = \bra{n^{(0)}} \hbar^{-1} \partial_{k_{x,y}} H^{(0)} \ket{m^{(0)}}$, and therefore stays real and unrenormalized even in the presence of complex self-energy interactions. Furthermore,
\begin{equation} \label{Eq:OMM_self_energy}
    \begin{split}
        \tilde{m}_{n}^{z}({\bf k},\omega) =
        &- \frac{1}{B_{z}} \braket{n | (H^{(1)} + \Sigma^{(1)}) | n} \\
        &- e \hbar \sum_{m \neq n}
        (E_{n}^{(0)} - E_{m}^{(0)}) 
        \frac{\mathrm{Im} (v_{x}^{(n, m)} v_{y}^{(m, n)})}{(\epsilon_{n}^{(0)} - \epsilon_{m}^{(0)})^{2}}
        \text{,}
    \end{split}
\end{equation}
is the complex, self-energy corrected OMM \cite{Sundaram1999, Xiao2005, Xia02010, THONHAUSER2011}, which reduces to the non-interacting, energy-independent OMM when $\Sigma_{n}^{(0)} \rightarrow -i 0^{+}$, but also whenever $\Sigma_n^{(0)}$ is a band-independent constant. 

Taking the imaginary part of Eq.~\eqref{Eq:GB_Trace} gives the first order correction to the spectral function
\begin{equation} \label{eq:SelfEnergySpecFunc}
    \begin{split}
        A^{(1)} &({\bf k}, \omega; B_z) =
        \frac{e B_z}{\hbar} \sum_{n} \Omega_{n}^{z} A_{n}^{(0)} \\
        + B_{z} &\sum_{n} \mathrm{Re} (Z_{n} \tilde{m}_{n}^{z}) \frac{\partial A_{n}^{(0)}}{\partial \omega} 
        - \frac{B_z}{\pi} \sum_{n} \mathrm{Im} (Z_{n} \tilde{m}_{n}^{z}) \frac{\partial \mathrm{Re} \, G_{n}^{(0)}}{\partial \omega}
        \text{,}
    \end{split}
\end{equation}
with $Z_{n} = (1-\partial_{\omega} \Sigma_{n}^{(0)})^{-1}$ the complex quasiparticle residue. Here $A_{n}^{(0)} = - \tfrac{1}{\pi} \mathrm{Im} \, G_{n}^{(0)}$ is the usual ARPES Lorentzian of half-width $|\mathrm{Im} \, \Sigma_{n}^{(0)}|$ centered on $\tilde{E}_{n}^{(0)} = \epsilon_{n}^{(0)} + \mathrm{Re} \, \Sigma_{n}^{(0)}$, and the remaining two lineshapes follow from it by differentiating $G_{n}^{(0)} = (\omega - E_{n}^{(0)})^{-1}$. The Berry curvature therefore appears in the spectral function with its own distinct lineshape, undistorted by the self-energy, while the OMM, which is complex when $\Sigma$ is finite, is split across the other two functions. 

To extract these quantities experimentally, consider the antisymmetrized photoemission difference spectrum $\Delta I ({\bf k}, \omega; B_z) = I({\bf k}, \omega; + B_{z}) - I({\bf k}, \omega; - B_{z})$, which by Eq.~\eqref{eq:Photoemission_intensity_1stOrder_Linear} is
\begin{equation}\label{eq:DeltaI}
    \begin{split}
        \Delta &I ({\bf k}, \omega; B_z) 
        = \frac{4 \pi B_{z}}{\hbar} |M^{(0)} ({\bf k})|^2 f(\omega) \sum_{n} \Biggl[
            \frac{e}{\hbar} \Omega_{n}^{z} A_{n}^{(0)} \\
            &+ \mathrm{Re}(Z_{n} \tilde{m}_{n}^{z}) \frac{\partial A_{n}^{(0)}}{\partial \omega} 
            - \frac{1}{\pi}\mathrm{Im} (Z_{n} \tilde{m}_{n}^{z})\frac{\partial \mathrm{Re} \, G_{n}^{(0)}}{\partial \omega}
        \Biggr]
        \text{.}
    \end{split}
\end{equation}
Antisymmetrizing in this way cancels the zero-field intensity together with every even power of $B_{z}$, so Eq.~\eqref{eq:DeltaI} is the full response up to corrections of $\mathcal{O} (B_{z}^{3})$. This expression therefore directly relates a generically measurable quantity to the Berry curvature and the OMM. 

The three lineshapes of $A_{n}^{(0)}$, $\partial_{\omega} \mathrm{Re} \, G_{n}^{(0)}$, and $\partial_{\omega} A_{n}^{(0)}$ carry distinct parities about $\tilde{E}_{n}^{(0)}$. Both $A_{n}^{(0)}$ and $\partial_{\omega} \mathrm{Re} \, G_{n}^{(0)}$ are even with respect to energy, and $\partial_{\omega} A_{n}^{(0)}$ is odd. $\partial_{\omega} \mathrm{Re} \, G_{n}^{(0)}$ also has the property that it integrates to zero over all energy. Thus taking the photoemission difference spectrum of Eq. \eqref{eq:DeltaI} and integrating over the energy window $\mathcal{W}_{n} = [\tilde{E}_{n}^{(0)} - W_{{\bf k}, n}, \tilde{E}_{n}^{(0)} + W_{{\bf k}, n}]$ gives 
\begin{equation} \label{eq:I_Berry}
     \frac{\hbar}{2 e B_z} \int_{\mathcal{W}_{n}} \mathrm{d} \omega \; 
     \Delta I({\bf k}, \omega; B_z)
     = \frac{2 \pi}{\hbar} |M^{(0)} ({\bf k}) |^{2} \; \Omega_{n}^{z} ({\bf k}),
\end{equation}
allowing a direct extraction of the Berry curvature. Furthermore, weighting the integrand by $(\omega - \tilde{E}_{n}^{(0)})$ flips the parity of the lineshapes, Hence,
\begin{equation} \label{eq:I_OMM}
    \begin{split}
        - \frac{1}{2 B_z} \int_{\mathcal{W}_{n}}  \mathrm{d} \omega \; 
        (\omega - \tilde{E}_{n}^{(0)} ({\bf k})) \, \Delta I({\bf k}, \omega; B_z) 
        \qquad \\
        = \frac{2 \pi}{\hbar} | M^{(0)} ({\bf k}) |^{2} \; \mathrm{Re} (Z_{n} \tilde{m}_{n}^{z} ({\bf k}))
        \text{,}
    \end{split}
\end{equation}
giving us access to the renormalized orbital magnetic moment. This analysis requires us to assume that $f (\omega)$, $Z_{n}$ and $\tilde{m}_{n}^{z}$ are constant across the integration limits, and that the window is sufficiently large to remove the contribution $\partial_{\omega} \mathrm{Re} \, G_{n}^{(0)}$, which will add a $\sim 1 / W_{{\bf k}, n}$ residual. An alternative method would be to fit the functional forms of the lineshapes directly. 

\begin{figure}
    \centering
    \includegraphics[width=\linewidth]{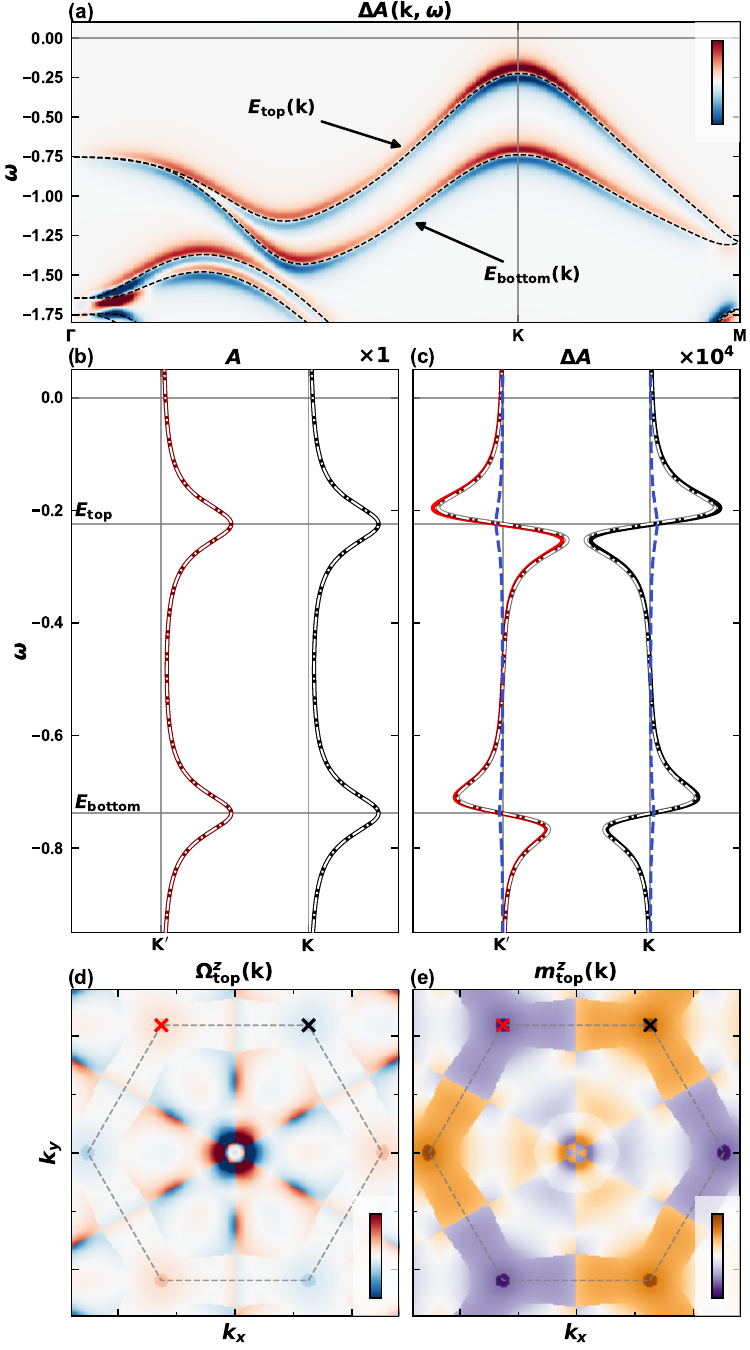}
    \caption{\textbf{Extracting Berry curvature and OMM using magneto-ARPES}:(a) Field-antisymmetrized spectral correction, $\Delta A = A(\mathbf{k},\omega;B_z)-A(\mathbf{k},\omega;-B_z) \propto \Delta I$, across the $\Gamma-{\bf  K}-{\bf M}$ path for a tight-binding model with spin-orbit coupling of WSe\textsubscript{2}, obtained directly from Ref.~\cite{Fang2015}. (b) Calculated energy distribution curve (EDC), $A(\mathbf{k},\omega)$, at $\mathbf{k} = {\bf K}$ (black) and $\mathbf{k} ={\bf K}^{\prime}$ (red) for the full spectral function, with solid and dashed lines showing the spectral function at positive and negative magnetic fields. (c) Field-antisymmetrized EDC, $\Delta A(\mathbf{k},\omega)$, at $\mathbf{k}={\bf K}, {\bf K}^{\prime}$, showing opposite behavior at ${\bf K}$ and ${\bf K}^{\prime}$. The signal has been amplified by a factor of $10^4$. Solid lines (black for ${\bf K}$ and red for ${\bf K}^{\prime}$) show $\Delta A$. White dot-dashed lines show the symmetric component about the center of $\Delta A$'s lineshape, and blue dashed lines its antisymmetric component. Integrating over the top band in (c) using Eqs.~\eqref{eq:I_Berry} and \eqref{eq:I_OMM} for all ${\bf k}$ gives (d) the Berry curvature $\Omega_\text{top}^{z} ({\bf k})$ and (e) the OMM $m_\text{top}^{z} ({\bf k})$ (which matches $\tilde{m}_\text{top}^z$ for constant $\Sigma$). Black and red crosses show the ${\bf K}$ and ${\bf K}^\prime$ points. For the data we have set $B_z=50$~mT and $\eta=50$~meV. The colorbars are relative to the maximal value, similar to the scales set in panel (c), and $\omega$ is in units of eV.}
    \label{fig:WSe2}
\end{figure}

As a concrete example, consider a monolayer of WSe\textsubscript{2}. This semiconducting material has been theorized to have strong Berry curvature for the spin-split band at the ${\bf K}$ and ${\bf K}^{\prime}$ points, with opposite sign in the different valleys \cite{Cho2018, Cho2021}. Using the two-dimensional tight-binding model with hoppings derived from density functional theory from Ref.~\cite{Fang2015}, assuming linearly polarized light, and taking a constant imaginary self-energy $\Sigma_n = \Sigma_{n}^{(0)} = -i \eta$, we find that the strongest signal in the field-antisymmetrized photoemission difference spectrum occurs indeed at the spin-split points ${\bf K}, {\bf K}'$, as well as for nearly degenerate bands below $-1$~eV at the $\Gamma$ point [Fig.~\ref{fig:WSe2}(a)]. 

In practice, to extract the Berry curvature and OMM for a given ${\bf k}$, we would measure an energy distribution curve (EDC) at both positive and negative magnetic field [Fig.~\ref{fig:WSe2}(b)] and subtract the two EDCs to obtain the field-antisymmetrized photoemission intensity [Fig.~\ref{fig:WSe2}(c)]. From this, we can already visually identify changes in sign by comparing the relative lineshapes, for example the opposite signs between the ${\bf K}$ and  ${\bf K}^{\prime}$ point. To directly extract the Berry curvature and renormalized OMM we then integrate this lineshape over the energy spanned by an individual band using either Eq.~\eqref{eq:I_Berry} or Eq.~\eqref{eq:I_OMM}. When repeated over all ${\bf k}$, we numerically obtain the full Berry curvature [Fig.~\ref{fig:WSe2}(d)] and OMM [Fig.~\ref{fig:WSe2}(e)] which we show for the top spin-split band highlighted by $E_{\text{top}} ({\bf k})$ in Fig.~\ref{fig:WSe2}(a). We have verified that these expressions exactly reproduce the conventional expressions calculated via Eq.~\eqref{eq:BerryCurvature} and Eq.~\eqref{Eq:OMM_self_energy}. 

We note that for this realistic example the detectable signal is expected to be weak, reflecting the weakness of the magnetic perturbation itself. For WSe\textsubscript{2} at a field of 50~mT (Fig.~\ref{fig:WSe2}), $\Delta I$ is approximately $10^4$ times smaller than the zero-field photoemission intensity. This ratio may be improved by the application of larger fields, or in materials with larger Berry curvature and OMM. However, in practice, measuring this signal will require careful normalization and may lie below the noise floor of current ARPES instrumentation. We also estimate that the change to the spectral function observed in the kagome material CsV\textsubscript{3}Sb\textsubscript{5} \cite{Huang2026} is several orders of magnitude larger than what we would predict here.

Our derivation of this first-order correction to the retarded Green's function under a magnetic field has deeper implications than the photoemission intensity alone: the same antisymmetrized velocity-Green's-function product structure of the itinerant part in Eq.~\eqref{eq:FirstOrderCorrection} also underlies the Kubo formula for the Hall conductivity \cite{TKNN1982, Xia02010, Crepieux2001, Ado2024}. This is not a coincidence: both are built from the same momentum-space curvature of the Bloch eigenstates. In our derivation, the antisymmetry between $\tilde{v}_{x}$ and $\tilde{v}_{y}$ arises from the Moyal star product encoding the non-commutativity of the kinetic momenta, \cite{Onoda2006Keldysh, Sugimoto2007GaugeCovariant}, whereas in the Kubo formula, the identical antisymmetry arises because a Hall response is the odd part of the conductivity tensor under exchange of current directions. Our correction can therefore be regarded as a momentum-resolved probe of the same geometric quantity that governs the intrinsic anomalous Hall conductivity, in the same sense that the spectral function is the momentum-resolved analog of the density of states. To further emphasize this, we note that for an insulating system, inserting our expression for the spectral function under the influence of a weak magnetic field into the well-known Streda formula \cite{Streda1982, Xiao2005} reproduces the quantized Hall conductivity
\begin{align}
    \sigma_{xy} &= e
    \frac{\partial N}{\partial B_z}  
    = e \int_{-\infty}^{E_{F}}  \mathrm{d} \omega \int_{\text{BZ}} \frac{\mathrm{d}^{2} {\textit{k}}}{(2 \pi)^2} \;
    \partial_{B_{z}} A^{(1)} ({\bf k}, \omega; B_z)  
\nonumber\\
    &= \frac{e^{2}}{h}\sum_{n \in \text{occ}} C_{n}
    \text{,}
\end{align}
where $N$ is the particle number and $C_{n} = \tfrac{1}{2 \pi} \int_{\text{BZ}} \mathrm{d}^{2} {\textit{k}} \;  \Omega_{n}^{z} ({\bf k})$ the Chern number of band $n$. 

In conclusion, we have presented a derivation of the ARPES-photoemission intensity expression in a weak magnetic field and found that it is directly related to the momentum-resolved Berry curvature and orbital magnetic moment. Despite the smallness of this signal, the result is exact to leading order in the field, and what stands between our findings and experimental observation is a question of experimental sensitivity rather than physical principle. Overcoming this limitation would enable the direct measurement of momentum-resolved Berry curvature and OMM with a single-particle probe for any quantum material.

\acknowledgements
We acknowledge insightful discussions with Henry Legg and Phil King. LCR acknowledges funding from EPSRC through UKRI3345. 

\bibliography{refrences}


\clearpage
\onecolumngrid

\setcounter{equation}{0}
\setcounter{figure}{0}
\setcounter{table}{0}
\setcounter{section}{0}
\setcounter{secnumdepth}{3}   
\renewcommand{\thefigure}{S\arabic{figure}}
\renewcommand{\thetable}{S\arabic{table}}
\renewcommand{\theequation}{S\arabic{equation}}
\renewcommand{\thesection}{S\arabic{section}}
\renewcommand{\theHfigure}{S\arabic{figure}}
\renewcommand{\theHtable}{S\arabic{table}}
\renewcommand{\theHequation}{S\arabic{equation}}
\renewcommand{\theHsection}{S\arabic{section}}

\setlength{\parindent}{0pt}
\setlength{\parskip}{6pt}

\begin{center}
\textbf{\large Supplemental Material for `Magneto-ARPES as a direct probe of Berry Curvature and the Orbital Magnetic Moment'}\\[0.5em]
Seohyun Kong, Bernd Braunecker and Luke C. Rhodes\\
\textit{SUPA, School of Physics and Astronomy, University of St Andrews, North Haugh, St Andrews, KY16 9SS, United Kingdom}
\end{center}

This Supplementary Material contains the full details of the calculations of the magnetic field corrections to the photoemission intensity. 
Section \ref{Supp:Moyal} introduces the fundamental definitions and properties of the Wigner transform, Weyl symbols and the magnetic Moyal star product.
In Section \ref{Supp:intensity} we derive the formal first order correction for the photoemission intensity and show that for linearly polarized light it depends only on the corrections of the spectral function. In Section \ref{Supp:GreensFunction} we compute the first order corrections for the Green's function. In Section \ref{Supp:SpecFunction} we connect the latter result with the spectral function and show how the Berry curvature and orbital magnetic moment arise from the result.

\section{Wigner Transform and the Magnetic Moyal Star Product} \label{Supp:Moyal}

In this section we construct the gauge-invariant Wigner transform and derive the magnetic Moyal star product on which the remainder of this supplement rests.

Magnetic field effects enter Hamiltonians and response functions through minimal coupling, which replaces the canonical momentum operator $\hat{\bkappa}$ by the kinetic momentum operator $\hat{\bf k} = \hat{\bkappa} + \frac{e}{\hbar} {\bf A} (\hat{\bf r})$ for an electron of charge $- e$, with $\hat{\bf r}$ being the position operator. The minimal coupling thereby introduces an explicit position dependence through the vector potential ${\bf A}$, which generates the field via $\nabla_{\bf r} \times {\bf A} ({\bf r}) = {\bf B} ({\bf r})$. This requires careful treatment: although a uniform magnetic field preserves the physical homogeneity of the system, no vector potential generating it can itself be invariant under ordinary spatial translations, so translational invariance is not manifest in any fixed gauge. This matters immediately for two-point objects like the real space Green's function. In a translationally invariant system, an operator kernel $\braket{{\bf r}_{2} | \hat{O} | {\bf r}_{1}}$ depends only on the separation ${\bf r}_{2} - {\bf r}_{1}$, so that a single Fourier transform carries it entirely into momentum space. Once ${\bf A} (\hat{\bf r})$ is present the kernel depends on ${\bf r}_{1}$ and ${\bf r}_{2}$ separately, and that reduction is no longer available. Nevertheless, one can still perform a coordinate substitution
\begin{equation} \label{Eq:Moyal01}
    {\bf R} = \frac{1}{2} ({\bf r}_{1} + {\bf r}_{2}) 
    \qquad \& \qquad
    {\bf r} = {\bf r}_{2} - {\bf r}_{1},
\end{equation}
which trades the two positions for a centre-of-mass coordinate ${\bf R}$ and a relative coordinate ${\bf r}$, with ${\bf r}_{1} = {\bf R} - \frac{1}{2} {\bf r}$ and ${\bf r}_{2} = {\bf R} + \frac{1}{2} {\bf r}$. Fourier transforming over the relative coordinate alone then gives a mixed real- and momentum-space representation of the operator, but such a transform inherits the gauge dependence of the kernel it acts on. Under a general gauge transformation ${\bf A} (\hat{\bf r}) \rightarrow {\bf A} (\hat{\bf r}) + \nabla_{{\bf r}} \chi (\hat{\bf r})$, the Hamiltonian, and hence every operator constructed from it, transforms as $\hat{O} \rightarrow \hat{U} \hat{O} \hat{U}^{\dagger}$ with $\hat{U} = e^{- i \tfrac{e}{\hbar} \chi (\hat{\bf r})}$, so that the kernel acquires a phase $\braket{{\bf r}_{2} | \hat{O} | {\bf r}_{1}} \rightarrow e^{- i \tfrac{e}{\hbar} (\chi ({\bf r}_{2}) - \chi ({\bf r}_{1}))}  \braket{{\bf r}_{2} | \hat{O} | {\bf r}_{1}}$, fixed entirely by the values of $\chi$ at the two endpoints.

Any quantity transforming with the same endpoint difference can therefore be used to compensate for the gauge dependence, and the line integral of the vector potential between the two points is precisely such a quantity, since a gradient integrates to the difference of its potential. Defining
\begin{equation} \label{Eq:Moyal02}
    \Phi ({\bf R}, {\bf r}) 
    = - \frac{e}{\hbar} \int_{{\bf R} - \frac{1}{2} {\bf r}}^{{\bf R} + \frac{1}{2} {\bf r}}
     \mathrm{d} \bm{\ell} \cdot {\bf A} (\bm{\ell})
    \text{,}
\end{equation}
taken along the straight line joining the endpoints, the same transformation shifts $\Phi \rightarrow  \Phi - \tfrac{e}{\hbar} (\chi ({\bf r}_{2}) - \chi ({\bf r}_{1}))$. Including this term as an additional exponential in the Fourier transform we can define the gauge-invariant Wigner transform  as
\begin{equation} \label{Eq:Moyal03}
    O_{W} ({\bf R}, {\bf k}) = \int \mathrm{d}^{d} {r} \; 
    e^{-i {\bf k} \cdot {\bf r}}
    e^{- i \Phi({\bf R}, {\bf r})}
    \braket{{\bf R} + \frac{1}{2} {\bf r} | \hat{O} | {\bf R} - \frac{1}{2} {\bf r}}
    \text{,}
\end{equation}
where $O_{W}$ is known as the Weyl symbol of the operator $\hat{O}$. The compensating factor, $e^{- i \Phi({\bf R}, {\bf r})}$, is a Wilson line. The resulting gauge-invariant phase-space variable is naturally identified with the kinetic wavevector ${\bf k}$, rather than the canonical wavevector $\bkappa$.

In this mixed representation the product of two operators is no longer the pointwise product of their Weyl symbols, since the phase-space variables do not commute. The Weyl symbol of the product of two operators then becomes the magnetic Moyal star product \cite{Moyal_1949,Hillery1984, Onoda2006Keldysh, Sugimoto2007GaugeCovariant}
\begin{align} \label{Eq:Moyal_StarProduct}
    (fg)_{W} ({\bf R}, {\bf k}) 
    &= f_{W}({\bf R}, {\bf k})\star g_{W}({\bf R}, {\bf k}) 
\nonumber\\
    &=
    f_{W} ({\bf R}, {\bf k})
    \exp \left( \frac{i}{2}
        \left[
            \overleftarrow{\nabla}_{\bf R} \cdot \overrightarrow{\nabla}_{\bf k} 
            - \overleftarrow{\nabla}_{\bf k} \cdot \overrightarrow{\nabla}_{\bf R} 
        \right]
        - \frac{i e B_{z}}{2 \hbar}
        \left[
            \overleftarrow{\nabla}_{\bf k} \times \overrightarrow{\nabla}_{\bf k} 
        \right]_{z}
    \right)
    g_{W} ({\bf R}, {\bf k}),
\end{align}
where the arrow direction indicates which Weyl symbol the derivatives act on. The first term in the exponent is the usual Moyal contribution from the non-commutativity of ${\bf R}$ and ${\bf k}$. The second is specific to the magnetic field and encodes $[\hat{k}_{i}, \hat{k}_{j}] = - i \tfrac{e}{\hbar} \, \varepsilon_{ijk} B_{k}$, the failure of the components of the kinetic momentum to commute with one another. We define 
\begin{align}
    &\star_0 = 
    \exp \left( \frac{i}{2}
        \left[
            \overleftarrow{\nabla}_{\bf R} \cdot \overrightarrow{\nabla}_{\bf k} 
            - \overleftarrow{\nabla}_{\bf k} \cdot \overrightarrow{\nabla}_{\bf R} 
        \right]
    \right),
 \label{S:Eq:star_zero}
 \\
    &\star_B =\exp \left(
    - \frac{i e B_{z}}{2 \hbar}
        \left[
            \overleftarrow{\nabla}_{\bf k} \times \overrightarrow{\nabla}_{\bf k} 
        \right]_{z}
    \right),
 \label{S:Eq:star_B}
\end{align}

to represent the two unique contributions with $\star = \star_0 \,\star_B$. 

These exponentials enable us to perform a Taylor expansion from which we can collect terms in powers of $B_z$. This is how we will perform a first-order expansion of the spectral function and photoemission intensity. 
If in addition both $f_W$ and $g_W$ are independent of $\bR$, then the product of two Weyl symbols simplifies to
\begin{equation} \label{Eq:Moyal05}
    (fg)_{W} ({\bf R}, {\bf k}) = f_{W}({\bf k})\star g_{W}({\bf k}) =
    f_{W}(\bk) \star_B g_W(\bk),
\end{equation}
as all $\nabla_\bR$ derivatives in $\star_0$ drop out.

Furthermore, we will make use of the following identities for the Weyl representation \cite{Hillery1984}.
First, the trace of an operator can be expressed in terms of an integral over its Weyl symbol
\cite{Hillery1984, Onoda2006Keldysh}
\begin{equation} \label{Eq:Trace_Weyl}
	\Tr[\hat{O}] = \int \frac{\dR\,\dk}{(2\pi)^d} \, \tr[O_W(\bR,\bk)]
	\text{,}
\end{equation}
where $\Tr$ is the full trace, $\tr$ the trace over the non-spatial degrees of freedom, and $d$ the spatial dimension.
Second, the trace of two operators can be expressed as
\begin{equation} \label{Eq:Trace_two}
	\Tr[\hat{f}\hat{g}]
	= \int \frac{\dR\,\dk}{(2\pi)^d} \, \tr[f_W \star g_W]
	= \int \frac{\dR\,\dk}{(2\pi)^d} \, \tr[f_W \, g_W]
	\text{,}
\end{equation}
 which is an integral over the conventional product of two Weyl symbols \cite{Hillery1984} and we have dropped the $\bR$ and $\bk$ dependences for brevity. If there are more than two operators, then, due to the associative properties of Weyl symbols, we can always tag one star product and consider its left and right factors as separate Weyl symbols following the rules of Eq.~\eqref{Eq:Moyal_StarProduct} such that, for instance,
\begin{equation}\label{Eq:Trace_many}
    \Tr[\hat{f}\hat{g}\hat{h}] 
	= \int \frac{\dR\,\dk}{(2\pi)^d} \tr\bigl[f_W \star g_W \star h_W\bigr] 
	= \int \frac{\dR\,\dk}{(2\pi)^d} \tr\bigl[f_W \, (gh)_W\bigr] 
    = \int \frac{\dR\,\dk}{(2\pi)^d} \tr\bigl[f_W \, (g_W\star h_W)\bigr].
\end{equation}
The latter equations will be convenient to simplify the expressions in the photoemission intensity.

\section{Derivation of the first-order correction to the photoemission intensity}
\label{Supp:intensity}

In this section we present the full derivation of the photoemission intensity written in terms of Weyl symbols and Moyal star products, which we will then use to obtain the first-order correction to the photoemission intensity under a weak magnetic field. 

We start from the standard expression for the photoemission intensity \cite{Damascelli2003,King2020}, obtained from Fermi's golden rule, retaining the full matrix structure and writing the trace explicitly as a sum over the internal degrees of freedom $\chi$ (for example the spin of the emitted electron) 
\begin{align}\label{eq:Photoemission_start}
	&I(\bk_f,\omega)
    \nonumber\\
    = &\frac{2\pi}{\hbar} \sum_{\chi,n,m}\int \dq_i
	\bra{\chi,\bk_f} \hat{M}(\hat{\bR},\hat{\bk}) \ket{n,\bq_i}
	\bra{n,\bq_i} \hat{A}(\omega) \ket{m,\bq_i}\bra{m,\bq_i}\hat{M}^\dagger(\hat{\bR},\hat{\bk})\ket{\chi,\bk_f} \,f(\omega) \,\delta(\bq_i-\bk_f+\bG).
\end{align}
Here, $\ket{n,\bq}$ is a Bloch basis 
and $\ket{\chi,\bk}$ the basis of the photoelectron. The labels $n,m$ are band indices, $\bk_f$ is the kinetic momentum of the measurable photoelectron, $\bq_i$ is the canonical crystal momentum for the Bloch states, $\bG$ is a reciprocal lattice vectors shifting $\bk_f$ into the first Brillouin zone if required, and $f(\omega)$ is the Fermi-Dirac distribution. The matrix element dipole operator is defined as
\begin{equation}
	\hat{M}(\hat{\bR},\hat{\bk}) = \frac{e}{m} \bA_{\text{ph}}(\hat{\bR}) \cdot \hat{\bk},
\end{equation}
with $\bA_{\text{ph}}(\hat{\bR})$ the photon vector potential,  $\hat{\bk} = \hat{\bkappa} + \frac{e}{\hbar} \bA(\hat{\bR})$ the kinetic momentum operator, and $\bA(\hat{\bR})$ the electromagnetic vector potential of the magnetic field. 

The idea now is to express this equation as a multiplication of operators, such that we can use Eq. \eqref{Eq:Trace_Weyl} to define the photoemission intensity in terms of a star-product of Weyl symbols, that we can subsequentially Taylor expand to obtain a first-order expansion.   

To do this we express the spectral function operator as 
\begin{equation}
\label{eq:A_operator}
	\hat{A}^{(n,m)}(\omega) = \int \dq_i \ \ket{n,\bq_i} \bra{n,\bq_i} \hat{A}(\omega) \ket{m,\bq_i} \bra{m,\bq_i}.
\end{equation}
This enables us to write 
\begin{equation}
	I(\bk_f,\omega) =\frac{2\pi}{\hbar} \sum_{\chi,n,m}
	\bra{\chi,\bk_f} \hat{M}(\hat{\bR},\hat{\bk})
	\hat{A}^{(n,m)}(\omega)
	\hat{M}^\dagger(\hat{\bR},\hat{\bk})
	\ket{\chi,\bk_f} f(\omega),
\end{equation}
which is a trace over the final state $\chi$ and $\bk_f$. Using the cyclic property of the trace, we can equivalently write  

\begin{equation} \label{Eq:Photoemission_operators}
	I(\bk_f,\omega) = \frac{2\pi}{\hbar} \sum_{\chi,n,m}
	\Tr\Bigl\{
	\hat{A}^{(n,m)}(\omega)
	\hat{M}^\dagger
	\hat{P}^{\chi,\bk_f}
	\hat{M}
	\Bigr\} f(\omega)
	\text{,}
\end{equation}
where we have introduced the
projector $\hat{P}^{\chi,\bk_f} = \ket{\chi,\bk_f}\bra{\chi,\bk_f}$. 

This is an unconventional way of writing the standard photoemission
intensity, but in this form it allows us to express Eq. \eqref{eq:Photoemission_start} as the trace of a product of operators. From Eq. \eqref{Eq:Trace_many} we can consequently write 
\begin{equation} \label{S:Eq:PE_Moyal_grouped}
	I(\bk_f,\omega) = \frac{2\pi}{\hbar} \sum_{\chi,n,m}
	\int \frac{\dR\,\dk}{(2\pi)^d} \tr
	\Bigl\{
	A^{(n,m)}_W \, \bigl( M^\dagger_W \star P^{\chi,\bk_f}_W \star M_W \bigr)
	\Bigr\} f(\omega)
	\text{,}
\end{equation}
where $\tr$ is the trace over the internal degrees of freedom (e.g. spin). Now the photoemission is an integral over the Weyl symbols of the spectral function operator ($A^{(n,m)}_W$), matrix element dipole operator ($M_W$) and photoelectron projector operator ($P^{\chi,\bk_f}_W$).

The star products remaining inside the bracket involve only the dipole
operator and the photoelectron projector. 
We show in detail in Sections~\ref{sec:M_W} and \ref{sec:P_W} that the Weyl symbols of these operators are
\begin{equation} \label{Eq:MW_full_0}
	M_W(\bR,\bk) = \frac{e\hbar}{m} \, \bA_\text{ph}(\bR) \cdot \bk
	\text{,}
\end{equation}
\begin{equation} \label{Eq:PW}
	P^{\chi,\bk_f}_W(\bR,\bk)
	=  \ket{\chi}\bra{\chi} \, \delta(\bk-\bk_f)
	\text{.}
\end{equation}
We see that $M_W(\bR,\bk)$ is the bare matrix element dipole operator with $\hat{\bR}$ replaced by $\bR$ and $\hat{\bk}$ replaced by the kinetic phase-space variable $\bk$. Thus $M_W$ carries no $B_z$ dependence. In the dipole approximation $\bA_\text{ph}$ also becomes $\bR$ independent, and we shall use 
\begin{equation} \label{Eq:MW_full_1}
	M_W(\bR,\bk) \approx M^{(0)}(\bk) = \frac{e\hbar}{m} \, \bA_\text{ph} \cdot \bk.
\end{equation}
Due to the independence of both $M^{(0)}$ and $P_W^{\chi,\bk_f}$ on $\bR$ it follows from Eq.~\eqref{Eq:Moyal05} that 
$M^{(0)\dagger} \star P_W^{\chi,\bk_f} \star M^{(0)} \ = M^{(0)\dagger} \star_B P_W^{\chi,\bk_f} \star_B M^{(0)}$. 
Performing then a Taylor expansion to first order in $B_z$ we find 
\begin{equation} \label{Eq:MPM}
	M^{(0)*} \star_B P^{\chi,\bk_f}_W \star_B M^{(0)}
	= \Bigr\{
        \bigl| M^{(0)} \bigr|^2 \ket{\chi}\bra{\chi}
	    - 
        \frac{e B_z}{\hbar} 
        \bigl[
		      \bigl( {\bf S} \times \nabla_\bk \bigr)_z 
		      + 
            \sigma_{\text{ph},z} 
	    \bigr]
        \ket{\chi}\bra{\chi}
    \Bigr\}
    \delta(\bk-\bk_f)
	+ \mathcal{O}(B_z^2)
	\text{,}
\end{equation}
with 
\begin{align} 
    {\bf S}(\bk)
    &= \mathrm{Im} \bigl[ M^{(0)*} \nabla_\bk M^{(0)} \bigr]
    = \Bigl( \frac{e \hbar}{m} \Bigr)^{2}
    \mathrm{Im} \bigl[ ( {\bf A}^{*}_{\text{ph}} \cdot \bk ) \, {\bf A}_{\text{ph}} \bigr]
    \text{,}
\label{Eq:S}
\\
    \sigma_{\text{ph},z} 
    &= \frac{1}{2} \mathrm{Im} \bigl[ \nabla_\bk M^{(0)} \times \nabla_\bk M^{(0)*} \bigr]_z
    = \frac{1}{2}\Bigr(\frac{e \hbar}{m} \Bigr)^{2} \,
    \mathrm{Im} \bigl[ {\bf A}_{\text{ph}} \times {\bf A}^{*}_{\text{ph}} \bigr]_z
    \text{.}
\label{Eq:sigma-phz}    
\end{align}

The spectral function is also modified by the the magnetic field, we therefore expand 
$A_W^{(n,m)} = A^{(0)\, (n,m)} + A^{(1)\, (n,m)} + \dots$, where $A^{(0)\, (n,m)}$ is the zero field limit and $A^{(1)\,(n,m)}$ is proportional to $B_z$. Retaining all terms to first order in $B_z$, the photoemission intensity becomes
\begin{align}\label{Eq:Intensity_full}
	I(\bk_f,\omega; B_{z}) =
    \frac{2\pi}{\hbar}
    &\sum_{\chi,n,m} \int \frac{\dR\,\dk}{(2\pi)^d} \tr
    \biggl\{ 
        |M^{(0)}|^2
        \bigl(A^{(0)\,(n,m)} + A^{(1)\, (n,m)} \bigr)\ket{\chi}\bra{\chi} \,
\nonumber\\
        &+ 
        \frac{e B_{z}}{\hbar} \Big(
            \big[ {\bf S} \times \nabla_{\bf k}
            \big( A^{(0)\, (n,m)} \big) \big]_{z}
            + 
            \sigma_{\text{ph}, z} \,A^{(0)\, (n,m)} 
        \Big)\ket{\chi}\bra{\chi}
    \biggr\}
    \delta(\bk-\bk_f)
    f(\omega)
    \text{.}  
\end{align}
In the zero field limit this formula leads to the conventional photoemission result,
\begin{align}
	I^{(0)}(\bk_f,\omega) 
    &=
    I(\bk_f,\omega;B_z=0)
    \nonumber\\
    &=
    \frac{2\pi}{\hbar}\sum_{\chi,n,m} \int \frac{\dR\,\dk}{(2\pi)^d} \tr
    \bigl\{ 
        |M^{(0)} (\bk)|^2
        A^{(0)\,(n,m)} (\bR, \bk, \omega) \ket{\chi}\bra{\chi} \, \delta(\bk-\bk_f)
    \bigr\} \
    f(\omega)
\nonumber\\
    &=
    \frac{2\pi}{\hbar} 
    |M^{(0)} (\bk_f)|^2
    \sum_{\chi,n,m} 
    \int \frac{\dR\,\dk}{(2\pi)^d} \tr
    \bigl\{ 
        A^{(0)\,(n,m)} (\bR, \bk, \omega) \ket{\chi}\bra{\chi} \, \delta(\bk-\bk_f)
    \bigr\} \
    f(\omega)
\nonumber\\
    &= \frac{2\pi}{\hbar}
    |M^{(0)}(\bk_f)|^2  
    \sum_{\chi,n,m} 
    \Tr\bigl[ \hat{A}^{(0) \, (n, m)} (\omega) \ \hat{P}^{\chi, \bk_f} \bigr] \
f(\omega)
\nonumber\\
    &= \frac{2\pi}{\hbar}
    |M^{(0)}(\bk_f)|^2  
    \sum_{\chi,n,m} 
    \bra{\chi, \bk_f} \hat{A}^{(0) \, (n,m)} (\omega) \ket{\chi, \bk_f} \
f(\omega)
\nonumber\\
    &= \frac{2\pi}{\hbar} 
|M^{(0)}(\bk_f)|^2  
    \tr \bigl[ \braket{ \bk_f | \hat{A}^{(0)} (\omega) | \bk_f} \bigr] \
f(\omega)
\nonumber\\
    &= \frac{2\pi}{\hbar} 
|M^{(0)}(\bk_f)|^2  
    A^{(0)}(\bq_i,\omega)\
f(\omega),
\label{eq:I0_final}
\end{align}
which is Eq.~(6) of the main text, where the matrix element is evaluated at
$\bk_f$ using the distribution identity $|M^{(0)}(\bk)|^2\delta(\bk-\bk_f)=|M^{(0)}(\bk_f)|^2\delta(\bk-\bk_f)$, leaving the integral over the product of two Weyl symbols, which is precisely the trace over the product of the operators $\hat{A}^{(0) \, (n, m)}$ and $\hat{P}^{\chi, \bk_f}$ using Eq.~\eqref{Eq:Trace_two}. The band sum is then carried out by completeness of the Bloch basis: from the definition in Eq.~\eqref{eq:A_operator}, since $\hat{A}^{(0)}$ is diagonal in the crystal momentum, the sum over $\chi$ becomes the trace over the internal degrees of freedom. As $\ket{\bk_f}$ overlaps only the crystal momentum $\bq_i = \bk_f - \bG$, with $\bG$ the unique reciprocal lattice vector folding $\bk_f$ into the first Brillouin zone, the last line is the spectral function as seen by the photoelectron. For the first order expression in $B_z$ we have
\begin{align}
	I^{(1)}(\bk_f,\omega; B_{z}) =
    \frac{2\pi}{\hbar}
    &\sum_{\chi,n,m} \int \frac{\dR\,\dk}{(2\pi)^d} \tr
    \biggl\{ 
        |M^{(0)}|^2
        A^{(1)\, (n,m)} \ket{\chi}\bra{\chi} \,
\nonumber\\
        &+ 
        \frac{e B_{z}}{\hbar} \Big(
            \big[ {\bf S} \times \nabla_{\bf k}
            \big( A^{(0)\, (n,m)} \big) \big]_{z}
            + 
            \sigma_{\text{ph}, z} \,A^{(0)\, (n,m)} 
        \Big)\ket{\chi}\bra{\chi}
    \biggr\}
    \delta(\bk-\bk_f)
    f(\omega)
    \text{.}  
\end{align}
Note, however, that $\bf S$ and $\sigma_{\text{ph}, z}$ from Eqs. \eqref{Eq:S} and \eqref{Eq:sigma-phz} are only non-zero if $\bA_\text{ph}$ is imaginary. This is true for circularly polarized light, but not true for linearly polarized light. Therefore, restricting our discussion to linearly polarized light we find
\begin{align}
	I^{(1)}(\bk_f,\omega,B_z) 
    &=
    \frac{2\pi}{\hbar}\sum_{\chi,n,m} \int \frac{\dR\,\dk}{(2\pi)^d} \tr
    \bigl\{ 
        |M^{(0)}|^2
        A^{(1)\,(n,m)} \ket{\chi}\bra{\chi} \,
    \bigr\}
    \delta(\bk-\bk_f)
    f(\omega)
\nonumber\\
    &= \frac{2\pi}{\hbar} 
|M^{(0)}(\bk_f)|^2  
    A^{(1)}(\bq_i,\omega)\
f(\omega),
\label{eq:I1_final}
\end{align}
which is Eq. (7) of the main text, again with $\bq_i = \bk_f - \bG$, following similar steps as with Eq.~\eqref{eq:I0_final}.

Thus the first order field-correction to the photoemission intensity under linearly polarized light is equivalent to the field-correction to the spectral function weighted by the zero-field matrix element effects already present in the measurement. We will evaluate $A^{(1)}$ by expanding the Dyson equation for $G_W$ to linear order in $B_z$ in Section \ref{Supp:SpecFunction}.

In the following subsections we provide the explicit Weyl symbols for $\hat{M}$ and $\hat{P}^{\chi,\bk_f}$.

\subsection{Weyl symbol for dipole matrix element $M_W$} 
\label{sec:M_W}
We start with the Weyl symbol of the dipole operator,
\begin{align}
	M_W(\bR,\bk)
	&= \int \dr \, \e^{-i \bk \cdot \br} \e^{-i \Phi(\bR,\br)}
	\bra{\bR+\tfrac{\br}{2}} \frac{e}{m}\bA_\text{ph}(\hat{\bR}) \cdot \hat{\bk} \ket{\bR-\tfrac{\br}{2}}
\nonumber\\
	&= \int \dr \, \e^{-i \bk \cdot \br} \e^{-i \Phi(\bR,\br)}
	\bra{\bR+\tfrac{\br}{2}} \frac{e}{m}\bA_\text{ph}(\hat{\bR}) \cdot \bigl[ \hat{\bkappa}+\frac{e}{\hbar}\bA(\hat{\bR}) \bigr] \ket{\bR-\tfrac{\br}{2}}.
\end{align}
For generality we keep here the $\hat{\bR}$ dependence of $\bA_\text{ph}(\hat{\bR})$.
The vector potential is diagonal in the position basis and can be applied to the left.
We start with splitting $M_W$ into two parts, $M_W = M_W^k + M_W^A$, with
\begin{align}
	M_W^k(\bR,\bk)
	&= \int \dr \, \e^{-i \bk \cdot \br} \e^{-i \Phi(\bR,\br)}
	\bra{\bR+\tfrac{\br}{2}} \frac{e}{m}\bA_\text{ph}(\hat{\bR}) \cdot \hat{\bkappa}\ket{\bR-\tfrac{\br}{2}},
\\
	M_W^A(\bR,\bk)
	&= \int \dr \, \e^{-i \bk \cdot \br} \e^{-i \Phi(\bR,\br)}
	\bra{\bR+\tfrac{\br}{2}} \frac{e}{m}\bA_\text{ph}(\hat{\bR}) \cdot \frac{e}{\hbar}\bA(\hat{\bR}) \ket{\bR-\tfrac{\br}{2}}.
\end{align}
The second equation yields
\begin{align}
	M_W^A(\bR,\bk)
	&= \int \dr \, \e^{-i \bk \cdot \br} \e^{-i \Phi(\bR,\br)}
	\frac{e}{m}\bA_\text{ph}(\bR+\tfrac{\br}{2}) \cdot \frac{e}{\hbar}\bA(\bR-\tfrac{\br}{2}) \bracket{\bR+\tfrac{\br}{2}}{\bR-\tfrac{\br}{2}}
\nonumber\\
	&= \int \dr \, \e^{-i \bk \cdot \br} \e^{-i \Phi(\bR,\br)}
	\frac{e}{m}\bA_\text{ph}(\bR+\tfrac{\br}{2}) \cdot \frac{e}{\hbar}\bA(\bR-\tfrac{\br}{2}) \, \delta(\br)
\nonumber\\
	&=
	\frac{e}{m}\bA_\text{ph}(\bR) \cdot \frac{e}{\hbar}\bA(\bR),
\end{align}
since $\Phi(\bR,0)=0$.

In the equation for $M_W^k$ the $\hat{\bkappa}$ operator could act directly as derivative, but that turns out to require a subtle
handling of a derivative of a delta function. Hence we choose to use first a momentum closure relation, using the variable $\bq$,
after which the handling of the delta function becomes straightforward. This leads to
\begin{align}
	M_W^k(\bR,\bk)
	&=
	\frac{e}{m}
	\int \frac{\dq}{(2\pi)^d}
	\int \dr \, \e^{-i \bk \cdot \br} \e^{-i \Phi(\bR,\br)}
	\bA_\text{ph}(\bR+\tfrac{\br}{2})
	\cdot
	\bq
	\
	\bracket{\bR+\tfrac{\br}{2}}{\bq}
	\bracket{\bq}{\bR-\tfrac{\br}{2}}.
\end{align}
The $\bq$ integration is over the true momentum, not crystal momentum. The wave functions are thus
plane wave states, such that
\begin{align}
	M_W^k(\bR,\bk)
	&=
	\frac{e}{m}
	\int \dr \, \e^{-i \bk \cdot \br} \e^{-i \Phi(\bR,\br)}
	\bA_\text{ph}(\bR+\tfrac{\br}{2})
	\cdot
	\int \frac{\dq}{(2\pi)^d}
	\bq
	\
	\e^{i \bq \cdot (\bR+\frac{\br}{2})}
	\e^{-i \bq \cdot (\bR-\frac{\br}{2})}
\nonumber\\
	&=
	\frac{e}{m}
	\int \dr \, \e^{-i \bk \cdot \br} \e^{-i \Phi(\bR,\br)}
	\bA_\text{ph}(\bR+\tfrac{\br}{2})
	\cdot
	(-i \nabla_\br)
	\int \frac{\dq}{(2\pi)^d}
	\e^{i \bq \cdot \br}
\nonumber\\
	&=
	\frac{e}{m}
	\int \dr \, \e^{-i \bk \cdot \br} \e^{-i \Phi(\bR,\br)}
	\bA_\text{ph}(\bR+\tfrac{\br}{2})
	\cdot
	(-i \nabla_\br)
	\delta(\br)
\nonumber\\
	&=
	\frac{e}{m}
	\int \dr \,
	\left[
	i \nabla_\br
	\cdot
	\left(
		\e^{-i \bk \cdot \br} \e^{-i \Phi(\bR,\br)}
		\bA_\text{ph}(\bR+\tfrac{\br}{2})
	\right)
	\right]
	\delta(\br)
\nonumber\\
	&=
	\frac{e}{m}
	\int \dr \,
	\left\{
	\left[
		\bk + \nabla_\br \Phi(\bR,\br)
	\right]
	\cdot
	\bA_\text{ph}(\bR+\tfrac{\br}{2})
	+
	i\nabla_\br
	\cdot
	\bA_\text{ph}(\bR+\tfrac{\br}{2})
	\right\}
	\e^{-i \bk \cdot \br} \e^{-i \Phi(\bR,\br)}
    \,
	\delta(\br)
\nonumber\\
	&=
	\frac{e}{m}
	\left\{
	\left[
		\bk + \nabla_\br \Phi(\bR,0)
	\right]
	\cdot
	\bA_\text{ph}(\bR)
	+
	\nabla_\br
	\cdot
	\bA_\text{ph}(\bR+\tfrac{\br}{2})
	\big|_{\br=0}
	\right\}
	\e^{-i \Phi(\bR,0)}.
\end{align}
To obtain the dipole operator it has to be assumed that $\bA_{\text{ph}}$ is in the Coulomb gauge and hence
$\nabla_\br \cdot \bA_{\text{ph}}=0$. Furthermore $\Phi(\bR,0)=0$ and
\begin{equation}
	\nabla_\br \Phi(\bR,\br)
	=
	-
	\frac{e}{\hbar} \nabla_\br
	\int_{\bR-\frac{\br}{2}}^{\bR+\frac{\br}{2}} \dd\bm{\ell} \cdot \bA(\bm{\ell})
	=
	-
	\frac{e}{\hbar}
	\left[
		\frac{1}{2}
		\bA(\bR+\tfrac{\br}{2})
		-
		\frac{-1}{2}
		\bA(\bR-\tfrac{\br}{2})
	\right]
	\overset{\br \to 0}{\longrightarrow}
	-
	\frac{e}{\hbar}
	\bA(\bR),
\end{equation}
which leads to the result
\begin{align}
	M_W^k(\bR,\bk)
	&=
	\frac{e}{m}
	\left[
		\bk - \frac{e}{\hbar}\bA(\bR)
	\right]
	\cdot
	\bA_\text{ph}(\bR).
\end{align}
Since we are dealing now with numbers and not operators the order of the factors does not matter, and we see that the $\bA$ dependent term is just the opposite to $M_W^A$.
Put together with $M_W^A$ we obtain
\begin{align}\label{Eq:MW_full}
	M_W(\bR,\bk) = 
	\frac{e}{m}
	\bA_\text{ph}(\bR)
	\cdot
	\bk. 
\end{align}
This shows also that the $\bk$ in the Weyl symbol represents the kinetic momentum and thus does not have an internal field dependence.

\subsection{Weyl symbol for projector $P^{\chi,\bk_f}_W$}
\label{sec:P_W}
We next focus on the projector $\hat{P}^{\chi,\bk_f} = \ket{\chi,\bk_f} \bra{\chi,\bk_f}$.
Here $\bk_f$ is the momentum of the outgoing photoelectron and is fixed.
Thus we wish to compute
\begin{equation}
	P^{\chi,\bk_f}_W(\bR,\bk)
	= \int \dr \, \e^{-i \bk \cdot \br} \e^{-i \Phi(\bR,\br)}
	\bracket{\bR+\tfrac{\br}{2}}{\chi,\bk_f} \bracket{\chi,\bk_f}{\bR-\tfrac{\br}{2}}.
\end{equation}
The subtle question here is what state $\bk_f$ describes. In the presence of the magnetic field, the eigenstates
of the free Hamiltonian no longer are simple plane waves but they acquire an extra phase factor similar to $\Phi$.
Indeed, we seek eigenstates for the kinetic momentum operator,
\begin{equation}
		\left[-i \nabla_\bR + \frac{e}{\hbar}\bA(\hat{\bR})\right] \psi(\bR,\bk_f)
		=
		\bk_f \psi(\bR,\bk_f),
\end{equation}
which are given by
\begin{equation}
	\psi(\bR,\bk_f) = \frac{1}{(2\pi)^{d/2}} \e^{i \bk_f \cdot \bR - i \frac{e}{\hbar}\int_{\bR_0}^\bR \dd\bm{\ell} \cdot \bA(\bm{\ell})},
\end{equation}
with $\bR_0$ some reference point.

The normalisation is chosen such that $\int \dR \psi^*(\bR,\bk_f) \psi(\bR,\bk_f') = \delta(\bk_f-\bk_f')$.
We then write $\psi(\bR,\bk_f) = \bracket{\bR}{\bk_f}$ and $\ket{\chi,\bk_f} = \ket{\chi} 
\ket{\bk_f}$, which leads to
\begin{align}
	P^{\chi,\bk_f}_W(\bR,\bk)
	&=
	\frac{\ket{\chi}\bra{\chi}}{(2\pi)^d}
	\int \dr \, \e^{-i \bk \cdot \br} \e^{-i \Phi(\bR,\br)}
	\e^{i \bk_f \cdot (\bR+\frac{\br}{2}) - i \frac{e}{\hbar}\int_{\bR_0}^{\bR+\frac{\br}{2}}\dd\bm{\ell} \cdot \bA(\bm{\ell})}
	\e^{-i \bk_f \cdot (\bR-\frac{\br}{2}) + i \frac{e}{\hbar}\int_{\bR_0}^{\bR-\frac{\br}{2}}\dd\bm{\ell} \cdot \bA(\bm{\ell})}
\nonumber\\
	&=
	\frac{\ket{\chi}\bra{\chi}}{(2\pi)^d}
	\int \dr \, \e^{-i \bk \cdot \br} \e^{-i \Phi(\bR,\br)}
	\e^{i \bk_f \cdot \br} \e^{- i \frac{e}{\hbar}\int_{\bR-\frac{\br}{2}}^{\bR+\frac{\br}{2}}\dd\bm{\ell} \cdot \bA(\bm{\ell})}.
\end{align}
We recognize in the last exponent $+i\Phi(\bR,\br)$ which cancels with the $-i\Phi(\bR,\br)$ from the Wigner
transformation. The remaining integral is a delta function,
\begin{align} \label{Eq:PW_full}
	P^{\chi,\bk_f}_W(\bR,\bk)
	&=
	\frac{\ket{\chi}\bra{\chi}}{(2\pi)^d}
	\int \dr \, \e^{-i (\bk-\bk_f) \cdot \br}
	=
	\ket{\chi}\bra{\chi}
	\delta(\bk-\bk_f).
\end{align}
Therefore $P_W^{\chi,\bk_f}$ is also independent of the magnetic field and $\bR$.

\section{Derivation of the first order correction to the Green's function} \label{Supp:GreensFunction}

We derive in this section the first-order correction to the Green's function in a weak magnetic field, retaining a general self-energy, which is required to obtain the correction to the spectral function.

We start by connecting the spectral function to the retarded Green's function. Noting that the traces in Eqs.~\eqref{eq:I0_final} and \eqref{eq:I1_final} can be written as
\begin{equation}
    \Tr[\hat{A}^{(j)}(\bq_i,\omega)] = \sum_n A^{(j) (n,n)}(\bq_i,\omega),
\end{equation}
for $j=0,1$, we can identify the matrix element of the spectral function with the imaginary part of the matrix element of the Green's function as
\begin{equation}
    A^{(j) (n,n)}(\bq_i,\omega) = -\frac{1}{\pi} \text{Im}\,G^{(j) (n,n)}(\bq_i,\omega).
\end{equation}
Here $G^{(0)}$ and $G^{(1)}$ are the zero-field and linear-$B_z$ parts of the Green's function 
\begin{equation}
    G^{n,m}(\bq_i,\omega) = \bra{n,\bq_i} [\omega_+ - \hat{H} -\hat{\Sigma}]^{-1}\ket{m,\bq_i},
\end{equation}
with $\omega_+ = \omega + i 0$.
Since the Bloch basis is natural for a crystalline structure, practical calculations will target $G^{n,m}$, whereas for the expansion in $B_z$ it will be more suitable to work with the Weyl symbols of the operator representation,
\begin{equation}
    \hat{G}(\omega) = \sum_{nm}\int \dq_i \ket{n,\bq_i} G^{(n,m)}(\bq_i,\omega) \bra{m,\bq_i} = [\omega_+ - \hat{H} - \hat{\Sigma}]^{-1}.
\end{equation}
Hence, after the expansion we shall switch back to the Bloch basis representation.

In terms of Weyl symbols, the Green's function is obtained through the equation
\begin{equation} \label{eq:Spectral01}
    \bigl[\omega_+ - H_W(\bR,\bk) - \Sigma_W(\bR,\bk,\omega)\bigr] \star G_W(\bR,\bk,\omega) = 1,
\end{equation}
for the Weyl symbols $H_W$ of the Hamiltonian and $\Sigma_W$ of the self-energy. 

To obtain the first-order correction in a weak field, we expand all Weyl symbols in Eq.~\eqref{eq:Spectral01} in powers of $B_{z}$, 
\begin{equation} \label{Eq:Spectral02}
    H_{W} = H^{(0)} + H^{(1)} + \dots\, ,
    \qquad
    \Sigma_{W} = \Sigma^{(0)} + \Sigma^{(1)} + \dots \, ,
    \qquad
    G_{W} = G^{(0)} + G^{(1)} + \dots \,
    \text{,}
\end{equation}
where the superscripts $(N)$ indicate proportionality to $B_z^N$. This expansion is done together with the expansion of the star product
\begin{equation}
    \star = \star_0 \star_B = \star_0 \left[1-\frac{ie B_z}{2\hbar}(\overleftarrow{\nabla}_{\bf k} \times \overrightarrow{\nabla}_{\bf k})_{z}+\dots\right].
\end{equation}
For the zero-field equation 
\begin{equation}
    \bigl[\omega_+ - H_W^{(0)}(\bR,\bk) - \Sigma_W^{(0)}(\bR,\bk,\omega)\bigr] \star_0 G_W^{(0)}(\bR,\bk,\omega) = 1,
\end{equation}
we note that due to translational invariance of the zero-field objects, all involved Weyl symbols are independent of $\bR$. As a consequence $\star_0$ reduces to the standard multiplication and the Weyl symbols reduce to the standard $\bk$ dependent matrix elements. The zero-field Green's function is thus computed from the usual momentum-resolved equation
\begin{equation}
    \bigl[\omega_+ - H^{(0)}(\bk) - \Sigma^{(0)}(\bk,\omega)\bigr] G^{(0)}(\bk,\omega) = 1,
\end{equation}
which is conventionally solved as a matrix equation in the Bloch basis. 
Collecting terms of order $\mathcal{O} (B_{z})$, we obtain
\begin{align} 
    \bigl[\omega_+ - H^{(0)}(\bk) - \Sigma^{(0)}(\bk,\omega)\bigr] G^{(1)}(\bk,\omega)
    - 
    \bigl[H^{(1)}(\bk) + \Sigma^{(1)}(\bk,\omega) \bigr] G^{(0)}(\bk,\omega)&
\nonumber\\
    + 
    \frac{i e B_{z}}{2 \hbar} 
    \Bigl\{\nabla_{\bf k} \bigl[H^{(0)}(\bk)+\Sigma^{(0)}(\bk,\omega) \bigr] \times \nabla_{\bf k} G^{(0)}(\bk,\omega)\Bigr\}_{z}
    &=0
    \text{,}
\label{Eq:Spectral03A}
\end{align}
where we can again identify the Weyl symbols with the regular $\bk$ dependent matrix elements.
Rearranging Eq.~\eqref{Eq:Spectral03A} for $G^{(1)}$, defining 
\begin{equation}
    \tilde{\bf v}(\bk,\omega) = \hbar^{-1} \nabla_{\bf k} \bigl[H^{(0)}(\bk) + \Sigma^{(0)}(\bk,\omega)\bigr]
\end{equation} as the renormalized band velocity, as well as utilizing the identity $\nabla_{\bf k} G^{(0)} = \hbar G^{(0)} \tilde{\bf v} G^{(0)}$ we obtain, 
dropping the labels $\bk,\omega$,
\begin{equation} \label{Eq:Spectral04}
    G^{(1)} =
    G^{(0)} \left[
    H^{(1)} + \Sigma^{(1)}
    + \frac{i e \hbar B_{z}}{2}
    (\tilde{v}_{y} G^{(0)} \tilde{v}_{x}
    - \tilde{v}_{x} G^{(0)} \tilde{v}_{y}) \right] G^{(0)}
    \text{.}
\end{equation}
This is the first-order correction to the Green's function. The term $H^{(1)} + \Sigma^{(1)}$ encodes the local orbital magnetic moment together with the Zeeman splitting, both renormalized by the self-energy. Since we consider only weak fields, it is sufficient to retain only the leading order correction. Stronger fields can be treated by carrying out the same expansion to $\mathcal{O} (B_{z}^{2})$ and beyond.

The $B_z$-dependent Hamiltonian $H^{(1)}$ follows from first order perturbation theory,
\begin{equation}
    H^{(1)}= \mu_{B} B_{z} (L_{z} + g_{s} S_{z})
    \text{,}
\end{equation}
where $\mu_{B}$ is the Bohr magneton, $g_{s}$ is the electron spin $g$-factor, and $L_{z}$ and $S_{z}$ define the orbital and spin angular momentum operators, encoding the local orbital magnetic moment and Zeeman splitting respectively. The form of $\Sigma^{(1)}$ depends on the type of self-energy correction considered; whatever that form, it acts as a correction to $H^{(1)}$, renormalizing both the orbital magnetic moment and the Zeeman splitting.

\section{Derivation of the first order correction to the Spectral function} \label{Supp:SpecFunction}

In this final section, we show that the imaginary part of the trace of the Green's function defined in \eqref{Eq:Spectral04} is directly connected to the Berry curvature of the bare band structure and to the self-energy renormalised orbital magnetic moment.

We start with the spectral representation of $G^{(0)}$,
\begin{equation} \label{Eq:Spectral05}
    G^{(0)}(\bk,\omega) = \sum_n \frac{\ket{n^{(0)},\bk}\bra{n^{(0)},\bk}}{\omega_+ - \epsilon_n^{(0)}(\bk) - \Sigma_n^{(0)}(\bk,\omega)}
    = \sum_n \frac{\ket{n^{(0)},\bk}\bra{n^{(0)},\bk}}{\omega_+ - E_n^{(0)}(\bk,\omega)},
\end{equation}
where we make the usual assumption that the self-energy is band diagonal, and choose $\ket{n^{(0)},\bk}$ and $\bra{n^{(0)},\bk}$ as a biorthonormal basis of eigenstates for $H^{(0)}(\bk)+\Sigma^{(0)}(\bk,\omega)$ with complex eigenvalues $E_n^{(0)}(\bk,\omega) = \epsilon_n^{(0)}(\bk) + \Sigma_n^{(0)}(\bk,\omega)$. The label $n$ remains the band index as long as the self-energy remains band diagonal, and then $\epsilon_n^{(0)}(\bk)$ is the corresponding Bloch band energy in the absence of the self-energy.
The trace over the imaginary part produces the spectral function contribution to the photoemission intensity,
\begin{align}
    A^{(0)} (\bk, \omega) 
    &=
    \Tr[ \hat{A}^{(0)}(\bk,\omega) ]
    =
    - \frac{1}{\pi} \mathrm{Im}\, \Tr[G^{(0)}(\bk,\omega)]
\nonumber\\
    &=
    - \frac{1}{\pi} \sum_{n}
    \frac{\mathrm{Im} \, \Sigma_{n}^{(0)} ({\bf k}, \omega)}{\bigl[\omega - \epsilon_{n}^{(0)} ({\bf k}) - \mathrm{Re} \, \Sigma_{n}^{(0)} ({\bf k}, \omega)\bigr]^{2} + \bigl[\mathrm{Im} \, \Sigma_{n}^{(0)} ({\bf k}, \omega)\bigr]^{2}}
    \text{.}
    \label{Eq:SpectralFunc_conventional}
\end{align}
The band-diagonal assumption of the self-energy is not required to compute Eq.~\eqref{Eq:Spectral04} numerically, but it is required in order to separate the ARPES photoemission intensity into band-dependent $\epsilon_n^{(0)}$ and self-energy $\Sigma_n^{(0)}$ expressions. The same assumption is also implicitly required \cite{Damascelli2003, King2020} when writing down the standard ARPES photoemission intensity in Eq. \eqref{Eq:SpectralFunc_conventional}.
A diagonalisation as in Eq.~\eqref{Eq:Spectral05} remains possible if the self-energy mixes bands, but then $n$ corresponds to a renormalized band index, and the eigenvalues $E_n^{(0)}$ no longer formally allow for a splitting between $\epsilon_n^{(0)}$ and $\Sigma_n^{(0)}$. We note however that it is possible to define the band off-diagonal self-energies (when $n\neq m$) as

\begin{equation}
\Sigma^{(n,m)}({\bf k}, \omega) = \frac{\bra{n^{(0)}, \bk}[\Sigma_{\text{orb}}({\bf k}, \omega),H_{\text{orb}}({\bf k})]\ket{m^{(0)}, \bk}}{\epsilon_m({\bf k}) - \epsilon_n({\bf k})}.
\end{equation}

If the self-energies are diagonal in the orbital basis ($a,b$) then this can be simplified to 
\begin{equation}
\Sigma^{(n,m)}({\bf k}, \omega) = \frac{\sum_{a,b} c^*_{an}c_{bm}(\Sigma_{\text{orb}}^{(a,a)}({\bf k}, \omega)-\Sigma_{\text{orb}}^{(b,b)}({\bf k}, \omega))H_{\text{orb}}^{(a,b)}({\bf k})}{\epsilon_m({\bf k}) - \epsilon_n({\bf k})},
\end{equation}
with $c_{an}$ the eigenvector connecting the orbital and band basis. This shows that the effect is dominant near band crossings. However, if the dominant orbital contributions to these bands have equal self-energies, e.g. by symmetry, the off-diagonal self-energy terms will also be zero. Additionally, if the overlap between the two states is zero via the Hamiltonian (i.e. the two dominant orbital components do not interact), the off-diagonal self-energies will also be small. In many cases, the band crossings of interest arise from bands with the same or similar self-energies, which is where our assumption is most strongly justified. It does however break down exactly at band crossings, and when considering strongly correlated electronic structures (e.g. Hubbard bands). We do not address such situations here.

Inserting Eq.~\eqref{Eq:Spectral05} into Eq.~\eqref{Eq:Spectral04} and taking the trace using $\mathrm{Tr} \, G^{(1)}(\bk,\omega) = \sum_{n} \braket{n^{(0)},\bk | G^{(1)}(\bk,\omega) | n^{(0)},\bk}$, we find
\begin{equation} \label{Eq:Spectral15}
    \mathrm{Tr} \, G^{(1)}
    = \sum_{n}
    \frac{\braket{n^{(0)}, \bk | (H^{(1)} + \Sigma^{(1)}) | n^{(0)}, \bk}}{(\omega_+ - E_{n}^{(0)})^{2}}
    + \frac{i e \hbar B_{z}}{2} \sum_{n, m}
    \frac{\tilde{v}_{y}^{(n, m)} \tilde{v}_{x}^{(m, n)} - \tilde{v}_{x}^{(n, m)} \tilde{v}_{y}^{(m, n)}}{(\omega_+ - E_{n}^{(0)})^{2}
    (\omega_+ - E_{m}^{(0)})}
    \text{,}
\end{equation}
where $\tilde{v}_{\mu}^{(n, m)} = \braket{n^{(0)}, \bk | \tilde{v}_{\mu} | m^{(0)}, \bk}$ for components $\mu=x,y,z$ of $\tilde{\mathbf{v}}$, and we omit henceforth the labels $\bk,\omega$ for better readability. 
We rewrite now the $\tilde{v}$ dependences in terms of the \emph{bare} band velocities, without self-energy corrections, defined by 
\begin{equation}
    {\bf v} = \hbar^{-1} \nabla_{\bf k} H^{(0)}.
\end{equation}
Then we have
\begin{align}
    \hbar v_\mu^{(n,m)} = (\epsilon_m^{(0)} -\epsilon_n^{(0)}) \braket{n^{(0)}, \bk|\partial_{k_\mu} m^{(0)}, \bk}  + \delta_{n,m} \partial_{k_\mu} \epsilon_m^{(0)},
\\
    \hbar \tilde{v}_\mu^{(n,m)} = (E_m^{(0)} -E_n^{(0)}) \braket{n^{(0)}, \bk|\partial_{k_\mu} m^{(0)}, \bk}  + \delta_{n,m} \partial_{k_\mu} E_m^{(0)},
\end{align}
and we can write the numerator of the second term in Eq.~\eqref{Eq:Spectral15} as

\begin{equation}
    \tilde{v}_{y}^{(n, m)} \tilde{v}_{x}^{(m, n)} - \tilde{v}_{x}^{(n, m)} \tilde{v}_{y}^{(m, n)} = 
    \begin{cases}
        - 
        2 i\frac{(E_{n}^{(0)} - E_{m}^{(0)})^{2}}{(\epsilon_{n}^{(0)} - \epsilon_{m}^{(0)})^{2}} 
        \mathrm{Im} (v_{x}^{(n, m)} v_{y}^{(m, n)})
         & \text{for $n\neq m$},
         \\
        0 & \text{for $n=m$},
    \end{cases}
\end{equation}
where the velocities on the right-hand side are now indeed the bare band velocities. We can therefore write 
\begin{equation} \label{Eq:Spectral15A}
    \mathrm{Tr} \, G^{(1)}
    = \sum_{n}
    \frac{\braket{n^{(0)}, \bk | (H^{(1)} + \Sigma^{(1)}) | n^{(0)}, \bk}}{(\omega_+ - E_{n}^{(0)})^{2}}
    + e \hbar B_{z} \sum_{n, m\neq n}
    \frac{(E_{n}^{(0)} - E_{m}^{(0)})^2}{(\epsilon_{n}^{(0)} - \epsilon_{m}^{(0)})^2}
    \frac{\mathrm{Im} (v_{x}^{(n, m)} v_{y}^{(m, n)})}{(\omega_+ - E_{n}^{(0)})^{2}
    (\omega_+ - E_{m}^{(0)})}
    \text{.}
\end{equation}

We now separate the poles of Eq.~\eqref{Eq:Spectral15A} via 
\begin{align}
    &\frac{\mathrm{Im} (v_{x}^{(n, m)} v_{y}^{(m, n)})}{(\omega_+ - E_{n}^{(0)})^{2}
    (\omega_+ - E_{m}^{(0)})} 
\nonumber\\
   &=
    \frac{\mathrm{Im} (v_{x}^{(n, m)} v_{y}^{(m, n)})}{(E_{n}^{(0)} - E_{m}^{(0)})(\omega_+-E^{(0)}_n)^2} 
    - 
    \frac{\mathrm{Im} (v_{x}^{(n, m)} v_{y}^{(m, n)})}{{(E_{n}^{(0)} - E_{m}^{(0)})^2(\omega_+-E^{(0)}_n})} 
    + 
    \frac{\mathrm{Im} (v_{x}^{(n, m)} v_{y}^{(m, n)})}{{(E_{n}^{(0)} - E_{m}^{(0)})^2(\omega_+-E^{(0)}_m})}
    \text{.}
\label{Eq:Seperate_poles}
\end{align}
The second and third terms can be shown to be equivalent, by swapping all $n$ and $m$ variables in the third term,
\begin{equation}
    \sum_{n, m \neq n} 
    \frac{\mathrm{Im} (v_{x}^{(n, m)} v_{y}^{(m, n)})}{{(E_{n}^{(0)} - E_{m}^{(0)})^{2} (\omega_+-E_{m}^{(0)}})} = 
    - \sum_{n, m\neq n}
    \frac{\mathrm{Im} (v_{x}^{(n, m)} v_{y}^{(m, n)})}{{(E_{n}^{(0)} - E_{m}^{(0)})^{2} (\omega_+-E_{n}^{(0)}})}
    \text{,}
\end{equation}
such that the trace of the Green's function with a band-diagonal self-energy becomes
\begin{equation} \label{Eq:TraceG1}
    \begin{split}
        \text{Tr} \, G^{(1)} 
        =
        &\sum_{n} 
        \frac{\braket{n^{(0)}, \bk | (H^{(1)} + \Sigma^{(1)}) | n^{(0)}, \bk}}{(\omega_+ - E_{n}^{(0)})^{2}} \\
        + e \hbar B_{z}
        &\sum_{n}
        \frac{1}{(\omega_+ - E_{n}^{(0)})^{2}}
        \sum_{m \neq n} 
        (E_{n}^{(0)} - E_{m}^{(0)}) \frac{\mathrm{Im} (v_{x}^{(n, m)} v_{y}^{(m, n)})}{(\epsilon_{n}^{(0)} - \epsilon_{m}^{(0)})^2} \\
        - 2 e \hbar B_{z}
        &\sum_{n}
        \frac{1}{\omega_+ - E_{n}^{(0)}}
        \sum_{m \neq n}
        \frac{\mathrm{Im} (v_{x}^{(n, m)} v_{y}^{(m, n)})}{(\epsilon_{n}^{(0)} - \epsilon_{m}^{(0)})^2}
        \text{,}
    \end{split}
\end{equation}
which is now written in terms of the bare, non-interacting band velocities. This is of central importance, as it allows the result to be related directly to the Berry connection of the band, which is defined by \cite{Berry1984}
\begin{equation}
    \mathcal{A}_{\mu}^{(n, m)} = i \braket{n^{(0)}, \bk | \partial_{k_{\mu}} m^{(0)}, \bk}.
\end{equation}
Using $\braket{n^{(0)}, \bk | \partial_{k_{\mu}} H^{(0)} | m^{(0)}, \bk} = \hbar v_{\mu}^{(n, m)} = \hbar \braket{n^{(0)}, \bk | v_{\mu} | m^{(0)}, \bk} = (\epsilon_{m}^{(0)} - \epsilon_{n}^{(0)}) \braket{n^{(0)}, \bk | \partial_{k_{\mu}} m^{(0)}, \bk}$, we see that the Berry curvature \cite{TKNN1982, Xia02010} connects to the velocities as
\begin{equation}
    \Omega_{n}^{z} 
    \equiv 
    - 2 \sum_{m \neq n} \mathrm{Im} (\mathcal{A}_{x}^{(n, m)} \mathcal{A}_{y}^{(m, n)})
    =
    - 2 \hbar^{2} \sum_{m \neq n}
    \frac{\mathrm{Im} (v_{x}^{(n, m)} v_{y}^{(m, n)})}{(\epsilon_{n}^{(0)} - \epsilon_{m}^{(0)})^2}
    \text{,}
\end{equation}
which is contained directly within the third term of Eq.~\eqref{Eq:TraceG1}. Furthermore, we define the renormalized orbital magnetic moment as
\begin{equation}
    \tilde{m}_{n}^{z} 
    \equiv
    - \frac{1}{B_{z}} \braket{n^{(0)}, \bk | (H^{(1)} + \Sigma^{(1)}) | n^{(0)}, \bk}
    - e \hbar
    \sum_{m \neq n} (E_{n}^{(0)} - E_{m}^{(0)})\frac{\mathrm{Im} (v_{x}^{(n, m)} v_{y}^{(m, n)})}{(\epsilon_{n}^{(0)} - \epsilon_{m}^{(0)})^2}
    \text{,}
\end{equation}
which contains both the local orbital magnetic moment $\braket{n^{(0)}, \bk | (H^{(1)} + \Sigma^{(1)}) | n^{(0)}, \bk}$ and an itinerant orbital magnetic moment contribution. This is not the conventional orbital magnetic moment, which is defined as \cite{Sundaram1999, Xiao2005, Xia02010,THONHAUSER2011}
\begin{equation}
    m_{n}^{z} 
    \equiv
    - \frac{1}{B_{z}} \braket{n^{(0)}, \bk |H^{(1)} | n^{(0)}, \bk}
    - e \hbar
    \sum_{m \neq n} \frac{\mathrm{Im} (v_{x}^{(n, m)} v_{y}^{(m, n)})}{\epsilon_{n}^{(0)} - \epsilon_{m}^{(0)}}
    \text{,}
\end{equation}
but it recovers the conventional orbital magnetic moment in the non-interacting limit $\Sigma \to -i 0^{+}$, or when $\Sigma$ can be assumed as a band-independent constant. Thus we refer to it as the \textit{renormalized} orbital magnetic moment. Taken together, this defines the trace of $G^{(1)}$ as
\begin{equation}
    \mathrm{Tr} \, G^{(1)} =
    B_{z}\frac{e}{\hbar} \sum_{n} \frac{\Omega_{n}^{z}}{\omega_+ - E_{n}^{(0)}}
    - B_{z} \sum_{n} \frac{\tilde{m}_{n}^{z}}{(\omega_+ - E_{n}^{(0)})^{2}}
    \text{.}
\end{equation}

Finally, the spectral function can then be obtained by taking the imaginary part of this trace. To connect with the bare spectral and Green's functions  
we note that with 
\begin{equation}
    G_n^{(0)} = \frac{1}{\omega_+ - E_n^{(0)}},    
\end{equation}
we have
\begin{equation}
    \frac{1}{(\omega_+-E_n^{(0)})^2} = [G_n^{(0)}]^2 = - Z_n \frac{\partial G_n^{(0)}}{\partial \omega },
\end{equation}
where the latter equality follows from 
\begin{equation} \label{Eq:Spectral22}
    - \frac{\partial G_{n}^{(0)}}{\partial \omega}
    = (1 - \partial_{\omega} \Sigma_{n}^{(0)}) [G_n^{(0)}]^2
    = Z_n^{-1} [G_n^{(0)}]^2
    \text{,}
\end{equation}
with the generalized, complex quasiparticle residue 
\begin{equation}
    Z_{n} \equiv (1 - \partial_{\omega} \Sigma_{n}^{(0)})^{-1}.
\end{equation} 
The product of a complex $Z_{n}$ with a complex derivative of the Green's function necessitates the calculation of both the real and the imaginary part of the Green's function. 

Defining the non-interacting spectral function $A_{n}^{(0)}$ through $\mathrm{Im} \, G_{n}^{(0)} = - \pi A_{n}^{(0)}$, our final expression becomes 
\begin{align} 
        A^{(1)}({\bf k}, \omega; B_z) 
        = 
        \Tr[\hat{A}^{(1)}(\bk,\omega)]
        =
        &\, 
        \frac{e B_z}{\hbar} \sum_{n} \Omega_{n}^{z} ({\bf k}) A_{n}^{(0)}({\bf k}, \omega) 
\nonumber\\
        + &B_{z} \sum_{n} \mathrm{Re} \bigl[ Z_{n} ({\bf k}, \omega) \, \tilde{m}_{n}^{z} ({\bf k}, \omega)\bigr] \frac{\partial A_{n}^{(0)}({\bf k}, \omega)}{\partial \omega}
\nonumber\\
        - &\frac{B_{z}}{\pi} \sum_{n} \mathrm{Im} \bigl[ Z_{n} ({\bf k}, \omega) \, \tilde{m}_{n}^{z} ({\bf k}, \omega)\bigr] \frac{\partial \mathrm{Re} \, G_{n}^{(0)}({\bf k}, \omega)}{\partial \omega}
        \text{,}
\label{S:eq:SelfEnergySpecFunc}
\end{align}
as quoted in Eq.~(15) in the main text. The three terms carry distinct lineshapes. The Berry curvature multiplies the Lorentzian $A_{n}^{(0)}$ itself, and so appears as a change in the spectral weight. The real part of the renormalized orbital magnetic moment multiplies $\partial_{\omega} A_{n}^{(0)}$, and so appears as a shift of the peak. The imaginary part, which is non-zero only for a complex self-energy, multiplies the dispersive lineshape $\partial_{\omega} \mathrm{Re} \, G_{n}^{(0)}$. These three shapes are linearly independent, which is what allows the Berry curvature and the orbital magnetic moment to be separated by fitting. 

Finally, if $\text{Im} \, \Sigma^{(0)}$ can be approximated as a band-independent constant, Eq.~\eqref{S:eq:SelfEnergySpecFunc} simplifies to
\begin{equation} \label{eq:SelfEnergySpecFunc_2}
    A^{(1)}({\bf k},\omega;B_z) =
    \frac{e B_z}{\hbar} \sum_{n} \Omega_{n}^{z} ({\bf k}) \, A_{n}^{(0)} ({\bf k}, \omega)
    + B_{z} \sum_{n} Z_{n} \, m_{n}^{z} ({\bf k}) \frac{\partial A^{(0)}_n}{\partial \omega} 
    \text{,}
\end{equation}
where $m_{n}^{z}$ is now the bare non-interacting orbital magnetic moment and $Z_n$ is the real quasiparticle residue, which itself is often approximated as an energy independent quasiparticle weight.

\end{document}